\documentstyle[twoside,fleqn,espcrc2]{article}

\def\vev#1{{\langle #1 \rangle}}

\def\half{{\textstyle{1\over2}}}
\def\){\right)} 
\def\({\left(} 
\def\]{\right]} 
\def\[{\left[}

\newcommand{\beq}{\begin{eqnarray}}
\newcommand{\eeq}{\end{eqnarray}}

\begin{document}

\title{
Non-perturbative improvement of bilinears in unquenched QCD 
\thanks{Supported by DOE contracts DE-FG03-96ER40956 and DOE-W7405-ENG-86.}
}

\author{T.~Bhattacharya\address{Los Alamos National Lab, MS B-285, Los Alamos,
                New Mexico 87545, USA},
R.~Gupta${}^{a}$,
W.~Lee${}^{a}$ and 
S.~Sharpe\address{Physics Department, Box 351560,
University of Washington, Seattle, WA 98195-1560, USA} 
\thanks{Speaker}}

\begin{abstract}
We describe how the improvement of quark bilinears generalizes from
quenched to unquenched QCD, and discuss which of the additional improvement
constants can be determined using Ward Identities.
\end{abstract}

\maketitle

A major motivation for undertaking the improvement program is to
facilitate unquenched calculations by allowing simulations at larger
lattice spacings. Thus it is important to study the application of
the improvement program to unquenched theories.
This has been done for the action itself,
and for certain operators in the chiral 
limit~\cite{alpha,JansenSommer}.
Here we take a further step
by considering the theory of on-shell improvement of
quark bilinears in unquenched QCD with non-zero quark masses.
We enumerate the additional improvement coefficients that
are required, and discuss which of them can be determined
non-perturbatively using  Ward Identities (WI).

The analysis depends on the number of dynamical flavors, and 
we consider here the physically relevant theories with
$N_f\ge 3$ non-degenerate flavors.
The two flavor theory is more complicated
and will be discussed elsewhere~\cite{Bhattinprep}.
We use an abbreviated notation for flavor traces
\beq
\vev{A_\mu} = {\sf Tr}(A_\mu) = \sum_{j=1,N_f} A_\mu^{(jj)},\ 
\vev{\lambda A_\mu} = {\sf Tr}(\lambda A_\mu),\nonumber
\eeq
with $\lambda$ an $SU(N_f)$ generator.
Note that we consider both flavor singlets and non-singlets,
with the latter being both off-diagonal and diagonal.
All three are needed for phenomenology. For example,
the flavor off-diagonal operator
$A_\mu^{(23)} = \bar d \gamma_\mu\gamma_5 s$ is needed to determine $f_K$,
and the nucleon matrix elements of the flavor diagonal operators
$A_\mu^{jj}$ and $T_{\mu\nu}^{jj}$ 
(which, with $j$ an unsummed flavor index, are linear combinations
of flavor singlet and non-singlet)
give information on the structure functions.

We begin by reviewing previous work on 
non-perturbative $O(a)$ improvement of unquenched QCD. 
The ALPHA collaboration~\cite{alpha}
has shown how on-shell improvement of the
action can be accomplished by adding the 
Sheikholeslami-Wohlert or ``clover'' term,
with appropriately chosen coefficient $c_{SW}$.
On-shell matrix elements of flavor non-singlet
axial, vector and tensor bilinears are then improved, 
in the chiral limit, by adding dimension four operators: 
\beq
\vev{\lambda A_\mu}^I &=& \vev{\lambda A_\mu} + 
c_A\, \partial_\mu \vev{\lambda P} , \\
\vev{\lambda V_\mu}^I &=& \vev{\lambda V_\mu} + 
c_V\, \partial_\nu \vev{\lambda T_{\mu\nu}} , \\
\vev{\lambda T_{\mu\nu}}^I &=& \vev{\lambda T_{\mu\nu}} + 
c_T [\partial_{\mu} \vev{\lambda V_{\nu}} \!-\! 
\partial_{\nu} \vev{\lambda V_{\mu}}] .
\eeq
Here we use the standard notation for local
lattice bilinears (see~\cite{alpha,Bhattetal}),
and, for brevity, set the lattice spacing $a$ to unity.
Flavor non-singlet
scalar and pseudoscalar bilinears are automatically on-shell $O(a)$
improved in the chiral limit.

The improvement coefficients $c_{SW}$ and $c_\Gamma$ ($\Gamma=V,A,T$)
depend on the number of dynamical quarks, $N_f$, and on the
bare coupling $g_0^2$. They can all be determined non-perturbatively
using WI.
In particular, enforcing the partial conservation of the 
improved flavor off-diagonal axial current determines
$c_{SW}$ and $c_A$, as well as the critical quark
mass $m_c$ at which chiral symmetry is restored~\cite{alpha}.
This method works for two or more flavors, and has been implemented
numerically for $N_f=2$~\cite{JansenSommer}.
With the improved axial current in hand, one can then enforce
the axial transformation properties of the vector and tensor
bilinears to determine $c_V$~\cite{GuagnelliSommer,Bhattetal} and 
$c_T$~\cite{Martinelli}. 

Away from the chiral limit there are many additional improvement
coefficients. In particular, the effective coupling constant 
becomes~\cite{alpha}
\beq
g_{\rm eff}^2 = g_0^2\, (1 + b_g \vev{M}/N_f) ,
\eeq
where $\vev{M}$ is the trace of the mass matrix.
Knowledge of the improvement coefficient $b_g$ allows
one to adjusts $g_0$, as $\vev{M}$ is varied, 
such that $g_{\rm eff}$ is constant.
This corresponds to working at fixed lattice spacing.
In this way one does not introduce spurious $O(a M)$ dependence
in physical quantities.
Methods for determining $b_g$ non-perturbatively are given in
Refs.~\cite{alpha,Martinelli}. 

This completes our review of previous work.
We now describe the new improvement coefficients
that arise when one improves bilinears
in the unquenched theory.
First we note that, in the chiral limit,
improvement of flavor singlet bilinears 
is more complicated than that of non-singlets,
due to the contributions from disconnected contractions.
For axial, vector and tensor bilinears the form remains the same, e.g.
\beq
\vev{A_\mu}^I = \vev{A_\mu} + \overline{c}_A \partial_\mu \vev{P},
\eeq
with new constants $\overline{c}_A$, $\overline{c}_V$ and $\overline{c}_T$. 
For the other bilinears there are additional gluonic operators\footnote{%
Normalized such that $S_{\rm glue}= (1/2 g_0^2) 
{\sf tr}(F_{\mu\nu}F_{\mu\nu})$.}
\beq
\vev{S}^I &=& \vev{S} + g_S\, {\sf tr}(F_{\mu\nu}F_{\mu\nu}) \\
\vev{P}^I &=& \vev{P} + g_P\, {\sf tr}(F_{\mu\nu}\tilde F_{\mu\nu}) 
\eeq
In principle, one can determine all five new improvement coefficients
by enforcing the invariance of these operators $O^I$ under
an off-diagonal vector transformation, i.e.
$ \delta_V^{(12)} O^I = 0 $.
This equation can be made non-trivial by taking appropriate
flavor non-singlet matrix elements.

Moving away from the chiral limit, there are several new improvement terms.
For flavor non-singlets, the general form is exemplified by
\beq
\lefteqn{\widehat\vev{\lambda A_\mu} = Z_A \left[ 
(1 + \overline{b}_A \vev{M}) \vev{\lambda A_\mu}^I\right.}\nonumber\\
&&\left.
+ b_A\half \vev{\{\lambda, M\} A_\mu} 
+ f_A \vev{\lambda M} \vev{A_\mu} \right] ,
\eeq
where the ``hat'' indicates an improved and renormalized operator.
CP invariance implies that only the anticommutator of 
$\lambda$ and $M$ appears.
It is instructive to consider examples of this general formula.
For flavor off-diagonal operators, the $f$-term drops out, leaving
\beq
\widehat{A_\mu^{(12)}} = Z_A (A_\mu^{(12)})^I
\left[1 + \overline{b}_A \vev{M} + b_A \half (m_1\!+\!m_2) \right]
\nonumber
\eeq
The $b_A$ term is that present in the quenched approximation,
while $\overline{b}_A$ multiplies the additional mass dependence arising
from quark loops.
All terms contribute for diagonal operators, e.g.
\beq
\lefteqn{\widehat{A_\mu^{(11)}}\!-\!\widehat{A_\mu^{(22)}} = Z_A 
\left[(A_\mu^{(11)}\!-\!A_\mu^{(22)})^I 
(1 + \overline{b}_A \vev{M})\right.}
\nonumber\\
&& \left. \!\!\!+ b_A (m_1 A_\mu^{(11)}\! -\! m_2 A_\mu^{(22)})
+ f_A (m_1\!-\!m_2) \vev{A_\mu} \right] 
\nonumber
\eeq
The $f$-term arises from disconnected contractions of the operator,
and is present only for non-degenerate quarks.

Finally, we consider the mass-dependent improvement coefficients needed
for flavor singlet operators. Here there are only two independent traces,
and thus two terms per bilinear, e.g.
\beq
\widehat\vev{A_\mu} = Z_A r_A\! \left[ 
(1 \!+\! \overline{d}_A \vev{M}) \vev{A_\mu}^I
\!\!+\! d_A \vev{ M A_\mu} \right]
\eeq
Note that the overall normalization constant differs from $Z_A$
by a factor $r_A$. For the axial current,
this factor is scale-dependent since the singlet current has a non-zero
anomalous dimension. For the other bilinears, however, $r_\Gamma$ is
a finite, scale-independent quantity, 
dependent only on the effective coupling constant and on $N_f$.
This is because the anomalous dimensions of the singlet and non-singlet
bilinears are the same.

In summary, the number of improvement terms
involving mass dependence increases substantially when one unquenches
and considers both singlet and non-singlet bilinears.
Improvement of flavor off-diagonal operators in
the quenched theory requires the determination of 5 such coefficients,
while complete improvement of bilinears in the unquenched theory 
requires 24.
In the quenched theory, all 5 coefficients can be determined using
WI with non-degenerate quarks~\cite{Bhattetal}.
We have studied the generalization of this method to unquenched
QCD, and find that all but three of the new coefficients can
be determined using WI with the following steps. 
\footnote{%
Some of combinations of coefficients depend on $N_f$,
and we quote results here for $N_f=3$.}
%

\noindent 1.
Enforcing the conservation of the diagonal vector charges 
while independently varying the three quark masses determines
all five coefficients associated with the vector bilinear
(as well as $r_V$).

\noindent 2.
Enforcing the PCAC relation away from the chiral limit
(combined with considerations outlined below)
determines the combinations $b_S$, $b_P-b_A$,
$\overline{b}_P-\overline{b}_A-\overline{b}_S$,
$f_S$, $d_S-3\overline{b}_S$, and
$\overline{d}_S-\overline{b}_S$, as well as $r_S$.
This generalizes the method of Ref.~\cite{Petronzio}.

\noindent 3.
Enforcing the off-diagonal axial transformation properties of operators,
for non-degenerate quark masses. This is the generalization of the
method of Ref.~\cite{Bhattetal}, and determines
$b_V+b_A$, $\overline{b}_V+\overline{b}_A$,
$b_P+b_S$, $2\sqrt{r_S}(\overline{b}_P-\overline{b}_S) + b_P-b_S$,
and $b_T$.

\noindent 4.
Enforcing off-diagonal vector transformation properties of operators, e.g.
\beq
\delta_V^{(12)} O^{(21)} = O^{(11)} - O^{(22)} ,
\quad
\delta_V^{(12)} \vev{O} = 0 .
\eeq
These identities have not been considered previously,
since they involve flavor diagonal bilinears.
They determine the $f_\Gamma$ ($\Gamma=V,A,T,S,P$)
together with $d_A$, $d_P$ and $d_T$, and $r_P$.

\noindent 5.
Enforcing the flavor off-diagonal axial transformations of
flavor singlet operators, e.g. $\delta_A^{(12)} \vev{P} = S^{(12)}$.
These determine $\overline{d}_P-\overline{b}_S$,
$\overline{d}_S-\overline{b}_P$, $\overline{d}_T-\overline{b}_T$,
and $r_T$.

The remaining undetermined combinations are
$\overline{b}_S\!+\!\overline{b}_P+\overline{d}_S\!+\!\overline{d}_P$,
$\overline{b}_T\!+\!\overline{d}_T$
and $\overline{d}_A$ (and $r_A$).

We close by expanding upon one feature of our analysis. In the second step,
when we enforce the PCAC relation, we need the $O(a)$ improved
expression for quark masses
\beq
\widehat{\vev{\lambda M}}\!=\!
Z_m \left[\vev{\lambda M}(1+\overline{b}_m \vev{M})
+ b_m \vev{\lambda M^2} \right] 
\eeq
\vspace{-0.15in}
\beq
\widehat{\vev{M}} \!=\!
Z_m r_m \left[\vev{M} (1+\overline{d}_m \vev{M})
+ d_m \vev{M^2} \right].
\eeq
Note that singlet and non-singlet masses are normalized differently,
so that individual masses have off-diagonal renormalization
even at $O(1)$:
\beq
\widehat{m}_j = Z_m \left[m_j + (r_m-1) \vev{M}/N_f + O(a) \right] .
\eeq
This effect is absent in the quenched approximation, 
where $r_m=1$, as well as $d_m=b_m$, $\overline{b}_m=\overline{d}_m=0$.

It turns out that these new improvement coefficients are related to
those we have previously introduced, and we need to determine these
relationships in order to extract all relevant information in step 2 above.
In other words, we need to generalize the quenched relations
$Z_m Z_S=1$ and $b_S=-2 b_m$ \cite{SintWeisz}.
To do so we note that it is consistent with
the WI to take the matrix elements of
the improved renormalized scalar density to be
\beq
\vev{H|\widehat{S^{(jj)}}|H} =
{\partial m_H \over \partial \widehat{m}_j}\bigg|_{\widehat{m}_{k\ne j}, a}
,
\eeq
with $H$ an arbitrary hadronic state. Note that the lattice spacing,
and not $g_0^2$, is held fixed in the derivative.
The r.h.s. can be evaluated in terms of derivatives known from
the form of the lattice action
\beq
{\partial m_H \over \partial m_j}\bigg|_{m_{k\ne j}, g_0} &=&
\vev{H|S^{(jj)}|H} ,\\
-2 g_0^4 {\partial m_H \over \partial g_0^2}\bigg|_{m_j}
&=& \vev{H|{\sf tr}(F_{\mu\nu}F_{\mu\nu})|H} .
\eeq
This results in the relations$^{\scriptstyle 4}$ $Z_S Z_m=1$,
$r_S r_m=1$, $b_S=- 2b_m$, $\overline{b}_S=-\overline{b}_m$,
$3 f_S = 2(b_m-d_m)$, and
$3\overline{d}_S = 3\overline{b}_m +2b_m-6\overline{d}_m-2 d_m$.
One also finds two constraints, namely
$d_S = 3 \overline{b}_S + b_S$, and
$2 g_S = b_g/g_0^2 $. The latter can be used as a consistency
check on the calculations of $g_S$ and $b_g$.



\begin{thebibliography}{9}
\bibitem{alpha} K.~Jansen {\em et al.}, Phys. Lett. B{372} (1996) 275;
M.~L\"uscher {\em et al.}, Nucl. Phys. B{478} (1996) 365.
\bibitem{JansenSommer} K.~Jansen and R.~Sommer, Nucl. Phys. B530 (1998) 185.
\bibitem{Bhattinprep} T.~Bhattacharya {\em et al}, in preparation.
\bibitem{Bhattetal} T.~Bhattacharya {\em et al}, hep-lat/9904011.
\bibitem{GuagnelliSommer} M.~Guagnelli and R.~Sommer,  
Nucl. Phys. B(Proc. Suppl.){63} (1998) 886.
\bibitem{Martinelli} G.~Martinelli {\em et al.}, Phys. Lett. B{411} (1997) 141.
\bibitem{Petronzio} G.M. de Divitiis {\em et al.}, 
Phys. Lett. B{419} (1998) 311.
\bibitem{SintWeisz} S.~Sint and P.~Weisz, 
Nucl. Phys. B(Proc. Suppl.){63} (1998) 856.

\end{thebibliography}
\end{document}